\documentclass{article}

\usepackage{graphicx}
\usepackage{epstopdf}
\usepackage{amsmath}         %% diverse Matheerweiterungen
\usepackage{amssymb}         %% diverse Matheerweiterungen, z.B. \mathbb{R}

\usepackage{tabularx}
\usepackage[utf8]{inputenc}

\title{Decomposition of formic acid}
\author{Martin Schmeißer  \\
	Chemnitz University of Technology  \\
	}

\date{\today}

\begin{document}

\maketitle

\begin{abstract}
Formic acid is known to act as a reduction agent for copper oxide. Its thermal uni-molecular decomposition was studied by means of DFT with special attention to reaction paths and kinetics.
\end{abstract}

\newpage

\section{Introduction} \label{Introduction}
Thermal decomposition of formic acid proceeds into the two possible products, \mbox{CO + H$_{2}$O} or \mbox{CO$_{2}$ + H$_{2}$}, of which CO is known to be responsible for reducing behaviour.\\
Columbia and Thiel provide an overview about the interaction of formic acid with transition metal surfaces \cite{Columbia1994}, including the respective reaction paths but nothing about reaction kinetics.
The thermal uni-molecular decomposition reactions have already been studied by Ko Saito \textit{et al.} \cite{Saito1984}. Their work includes ab-initio calculations at the Hartree-Fock (HF) and Møller-Plesset Perturbation Theory (MP-2) level and a quite exhaustive list of referenced experimental work as well as their own experiments of Ar diluted HCOOH in a shock tube system.\\
DFT calculations were previously performed by Jan W. Andzelm \textit{et al.} \cite{Andzelm1995}, CCSDT-1 calculations by Goddard \textit{et al.} \cite{Goddard1992}.\\
This work aims at simulating the reaction pathway of the decomposition of formic acid in the gas phase in order to learn about reaction energies and kinetics, but also to make predictions about reasonable process parameters for CuO reduction, and possibly to support the results of earlier work. It is meant as a starting point for a more thorough investigation of the reaction mechanisms of the reduction itself.

\section{Simulation Details} \label{simdetails}
The calculations of reaction energies were run with DMol$^{3}$ \cite{Delley1990} in Materials Studio and Turbomole \cite{Turbomole}.
Reaction paths were approximated with Synchronous Transit methods \cite{Halgren1977} as implemented in DMol$^{3}$, the guessed transition states refined and verified using Eigenvector Following methods as implemented in DMol$^{3}$ and Turbomole. The parameters for the DFT calculations were set as follows. For DMol$^{3}$ the DNP basis set, PBE \cite{Perdew1996c} functional and the default 'fine' settings were employed, i.e. SCF cycles were converged to an (RMS) energy change of $1 \cdot 10^{-6}$\,eV, geometries optimized to a maximum force of $2 \cdot 10^{-3}$\,Ha/\AA{} and a maximum displacement of $5 \cdot 10^{-3}$\,\AA{}. In Turbomole, the TZP basis set, b3-lyp \cite{Dirac1929,Slater1951,Vosko1980,Becke1988a,Lee1988,Becke1993} hybrid functional and equal convergence parameters were used.

\section{Reaction Energies} \label{Reaction Energies}
\begin{figure}
\includegraphics[width=.8\textwidth]{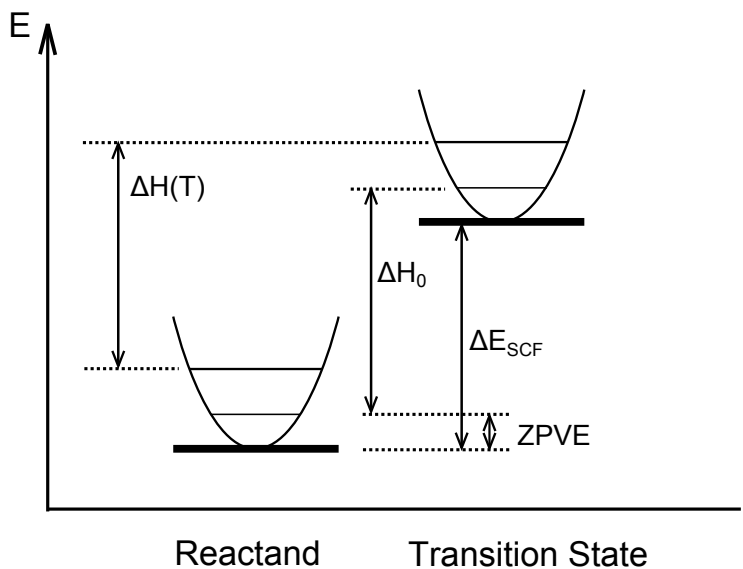}
\caption{Definition of the various Energy and Enthalpy Terms}
\label{quantities}
\end{figure}

\begin{figure}
\includegraphics[width=1.2\textwidth]{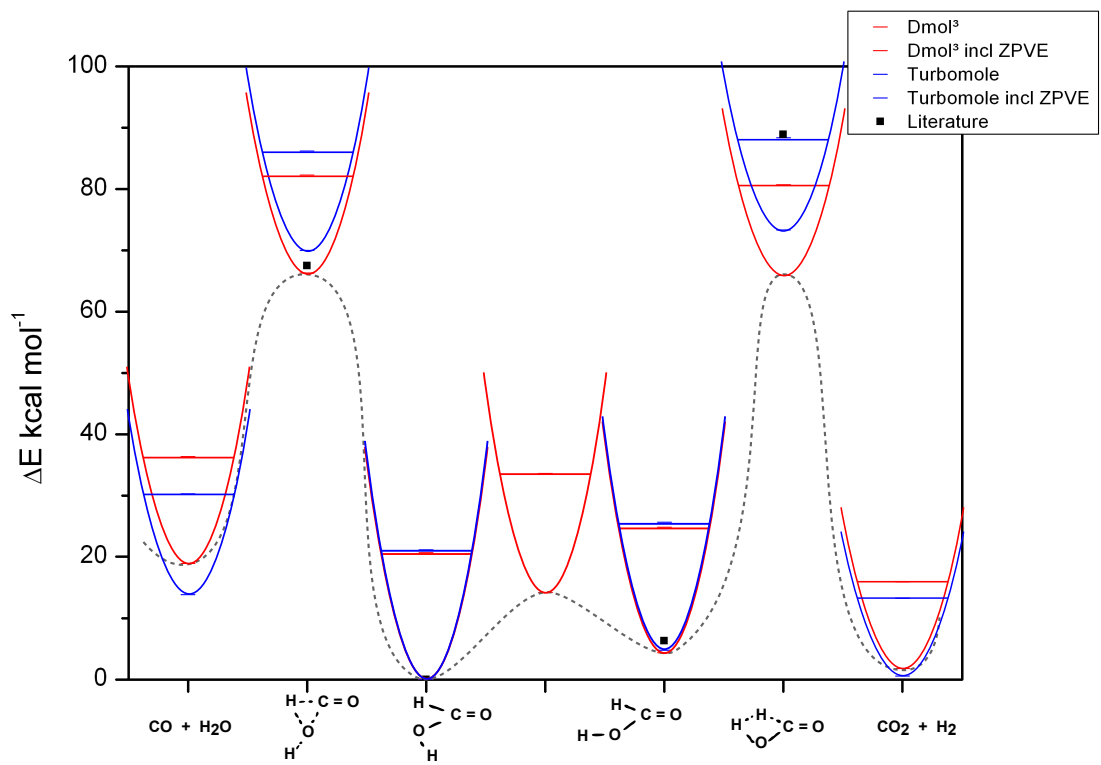}\\
\caption{DFT energies of the investigated structures. Literature values (black) from Ko Saito et al. \cite{Saito1984} do not include ZPVE. The dashed line illustrates a possible reaction path, the parabolas illustrate a harmonic approximation of vibrational energies, their slope is arbitrary.}
\label{energies}
\end{figure}

SCF Energies of the examined structures with corresponding zero point vibrational energies (ZPVE) are displayed in Figure \ref{energies}, activation energies are listed in Table \ref{actEnergies}. See Figure \ref{quantities} for an illustration of the various energy and enthalpy terms.\\
The reaction barriers for the two channels are nearly equal (DMol$^3$ predicts a slightly lower barrier for path 2, Turbomole a slightly higher one, but the differences are in the order of expected errors).\\
Previous \textit{ab-initio} calculations and measurements are discordant as to whether the reaction barriers are similar (Hsu \textit{et al.} (1982) \cite{Hsu1982}, Andzelm \textit{et al.} (1995) \cite{Andzelm1995} and  Goddard \textit{et al.} \cite{Goddard1992}) or significantly different (Saito et al. (1984)\cite{Saito1984} and Blake \textit{et al.} (1971) \cite{Blake1971}). The latter two predict a significantly higher (about 20 kcal mol$^-1$) barrier for path 2, and make this responsible for the higher conversion to \mbox{CO + H$_{2}$O}. However, looking at the current data it seems the difference between the reaction paths is rather due to thermodynamic factors, see section \ref{Kinetics} for explanation.

\begin{table}
\begin{tabularx}{0.8\textwidth}{|l|X|X|X|X|} \hline
 & \multicolumn{2}{c|}{\textbf{ $\Delta E_{SCF}$ }}  & \multicolumn{2}{c|}{\textbf{ $\Delta H_0$ }}\\
\hline
\textbf{Path} & DMol$^3$ & Turbomole & DMol$^3$ & Turbomole\\
\hline
\textbf{1} & 66 & 69 & 61 & 65 \\
\hline
\textbf{2} & 65 & 73 & 60 & 67 \\
\hline
\end{tabularx}
\caption{Energy differences between cis- and trans-HCOOH and their corresponding TS in kcal mol$^{-1}$}
\label{actEnergies}
\end{table}

\newpage
\section{Kinetics} \label{Kinetics}

\subsection{Theory}

\paragraph{}
Reaction kinetics are evaluated in terms of transition state theory, which is expected to give a good approximation in the case of thermal equilibrium (i.e. fast energy exchange of a molecule with the surrounding system). A more complex theory like RRKM is not employed because the large uncertainties in DFT calculations would still render the results inaccurate. It must be noted here, that transition state theory gives an upper bound to the rate constants.\\
The equilibrium constant for the first part of the reaction $K^{\ddagger}$ (from formic acid to the transition state) is
$$K^{\ddagger} = \frac{c(X^{\ddagger})}{c(HCOOH)} = \frac{z^{\ddagger}}{z_{HCOOH}}e^{-\Delta E_0 / R T}$$
where $c^{\ddagger}, c_{HCOOH}$ are the concentrations of the transition state (TS) and formic acid, $z^{\ddagger}, z_{HCOOH}$ their partition sums, respectively, and $\Delta E_0$ is the difference of ground state energies between reactant and transition state. R is the ideal gas constant. $z^{\ddagger}$ may now be written as
$$z^{\ddagger}=z_{\ddagger}\frac{1}{1-e^{-h\nu / k_B T}}$$
where $\nu$ is the frequency of the vibrational mode of the TS along the reaction coordinate and $z_{\ddagger}$ is the partition sum without that vibration. Since the corresponding force constant is very low, the frequency will also be very low and the exponential function is developed into a series and all but the linear term are neglected
$$z^{\ddagger}=z_{\ddagger}\frac{1}{1-e^{-h\nu / k_B T}} \approx z_{\ddagger} \frac{1}{1-(1-h\nu / k_B T)} = z_{\ddagger} \frac{k_B T}{h \nu}$$
The TS has $2\nu$ chances per unit time to dissociate ($\nu$ in each direction). Due to the very low force constant, one may assume, that every chance for dissociation will be used. However, only half of the chances will lead to the reaction products, the others will lead back to the reactants. Thus, the rate constant is
\begin{align*}
k &= 2 \nu \cdot \frac{1}{2} \cdot K^{\ddagger} \\
&= \nu \cdot z_{\ddagger} \cdot \frac{k_B T}{h \nu} \cdot \frac{1}{z_{HCOOH}} \cdot e^{-\Delta E_0 / R T}\\
&= \frac{k_B T}{h} \cdot \frac{z_{\ddagger}}{z_{HCOOH}} \cdot e^{-\Delta E_0 / R T}\\
&= \frac{k_B T}{h} \cdot e^{\Delta S(T) / R} \cdot e^{-\Delta H_{vib}(T) / R T} \cdot e^{-\Delta E_{SCF} / R T}
\end{align*}
which is known as the Eyring Equation. Here, S is the entropy, $H_{vib}$ the enthalpy of the system due to vibrations (including ZPVE) and $E_{SCF}$ the approximated ground state energy of the system. The temperature dependence of $\Delta S$ and $\Delta H$ is displayed in figures \ref{enthalpy1} and \ref{enthalpy2}. $\Delta H_{vib}$ is defined as
$$\Delta H_{vib} \equiv \Delta H - \Delta E_{SCF}$$

\subsection{Results and Discussion}

Reaction enthalpies, vibrational spectra and molecular structures were found to agree well between the two programs, therefore all of the presented kinetic results are from DMol$^{3}$ output.\\
The reaction rate constants are equal at about $280\,^{\circ}\mathrm{C}$
, at lower temperatures the decarboxylation reaction is preferred, at higher temperatures the dehydration reaction is.\\
Thermodynamic contibutions have a dramatic effect on the reaction rates. 
Neglecting all thermodynamic quantities but ZPVE in the Eyring equation yields a much higher (1-2 orders of magnitude) rate constant for path 2 compared to path 1. 
The following figures (\ref{enthalpy1}, \ref{enthalpy2} and \ref{ratecoeff}) display the temperature dependence of the two reaction enthalpy and entropy barriers and the rate coefficients.\\
Reaction barriers are consistent with other \textit{ab-initio} works, the reaction rates are consistent with experimental results from Blake \textit{et al.}  but several orders of magnitude smaller than the results from shock-wave experiments.\\
In order to exemplify these results we apply them to reactor conditions as described by Thomas Wächtler \cite{Wachtler2009}. The reduction is carried out at 388\,K ($115\,^{\circ}\mathrm{C}$) and 1.3\,mbar with a flow rate of 70\,mg per minute for formic acid and 180\,mg per minute for Ar as a carrier gas the formic acid gas will have a molar concentration of about $1 \cdot 10^{-2} \mathrm{\frac{mol}{m^3}}$. The rate constant for the dehydration reaction at $115\,^{\circ}\mathrm{C}$ is $2.42 \cdot 10^{-21} \, \mathrm{s^{-1}}$. Under these conditions the formic acid has about 0.4\,seconds to decompose before it reaches the sample (assuming one litre of gas volume between the supply and the sample). After that time the CO concentration in the gas will be $9.86 \cdot 10^{-24} \, \mathrm{\frac{mol}{m^3}}$. A copper oxide film that consists half of Cu$_{2}$O and half of CuO and measures 10\,nm $\cdot$ 1\,cm $\cdot$ 1\,cm contains about $6 \cdot 10^{-8} \, \mathrm{mol}$ oxygen. Therefore it would take $4 \cdot 10^{16} \, \mathrm{min}$ just to flow by the stoichiometrically equivalent amount of CO. See Appendix \ref{reactor} for the details of this calculation.\\
Clearly, the thermal decomposition cannot be the only reaction mechanism for the reduction.

\begin{figure}
\includegraphics[width=.8\textwidth]{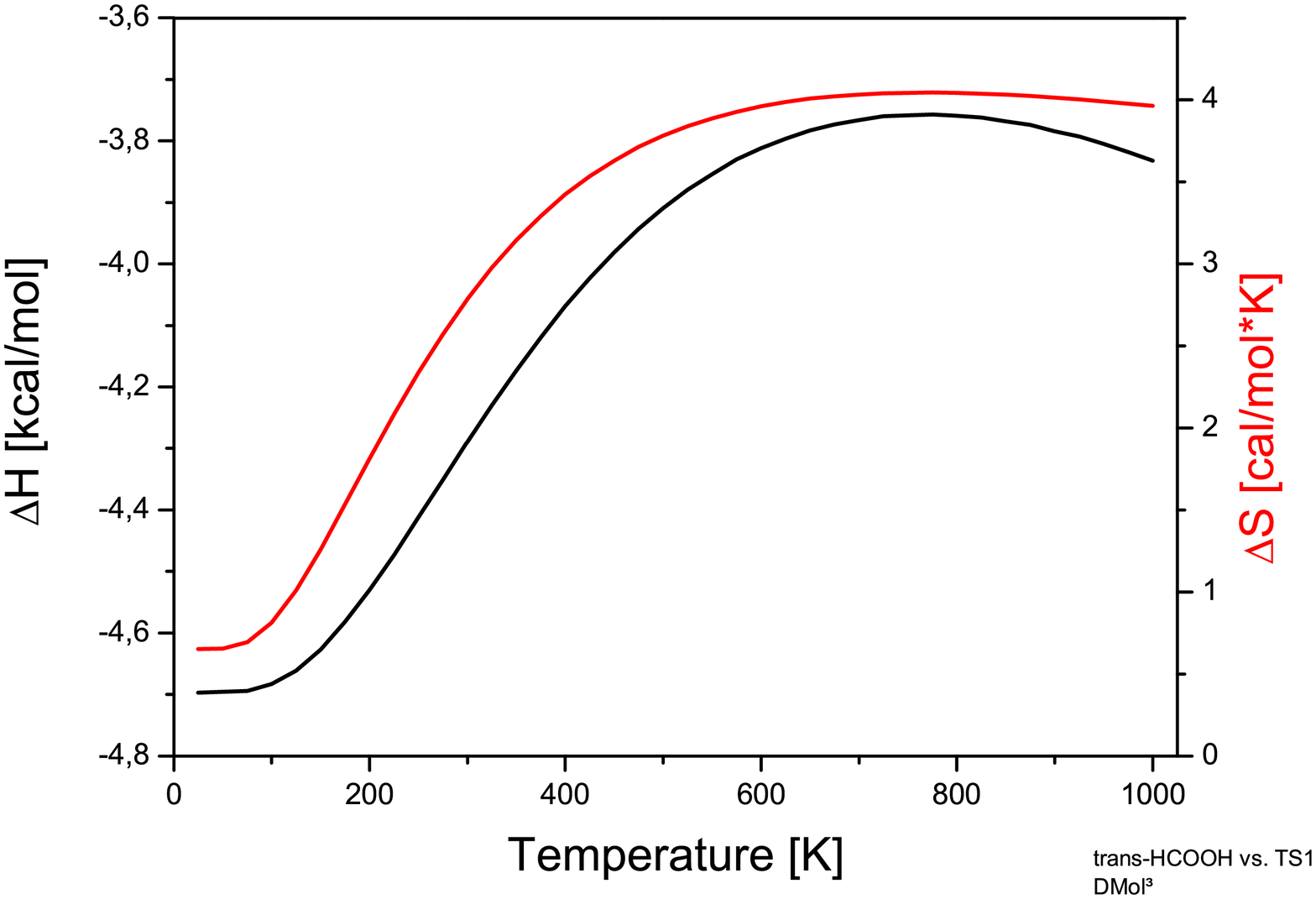}
\caption{Enthalpy and Entropy differences between trans-HCOOH and TS1 as function of temperature}
\label{enthalpy1}
\end{figure}

\begin{figure}
\includegraphics[width=.8\textwidth]{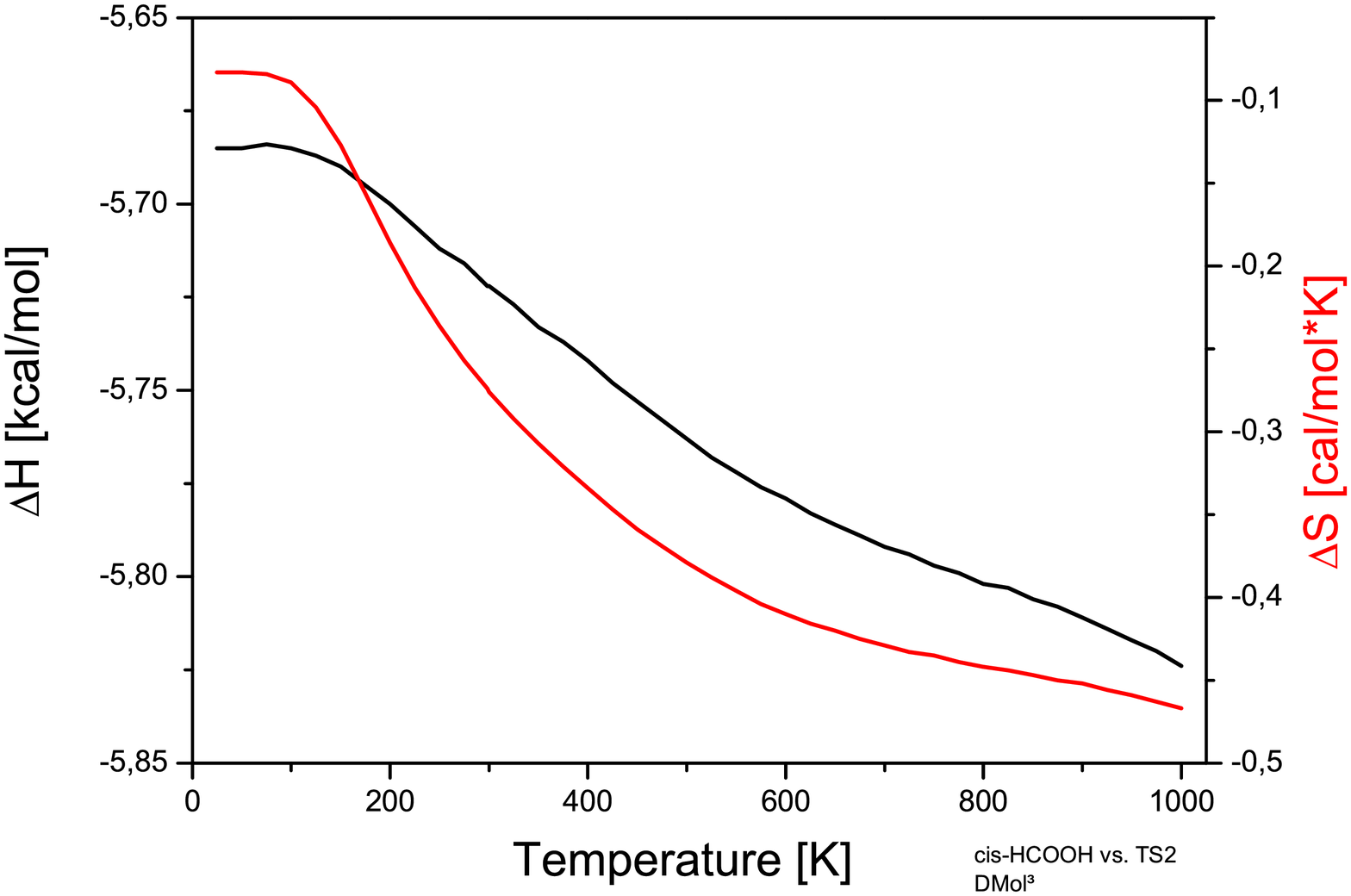}
\caption{Enthalpy and Entropy differences between cis-HCOOH and TS2 as function of temperature}
\label{enthalpy2}
\end{figure}

\begin{figure}
\includegraphics[width=.8\textwidth]{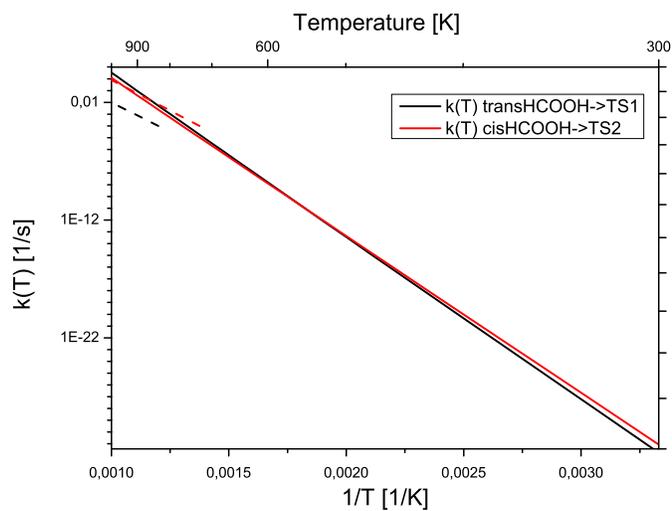}
\caption{Rate coefficients as function of temperature. Dashed lines are values from Blake \textit{et al.} 1971}
\label{ratecoeff}
\end{figure}

\begin{figure}
\includegraphics[width=.8\textwidth]{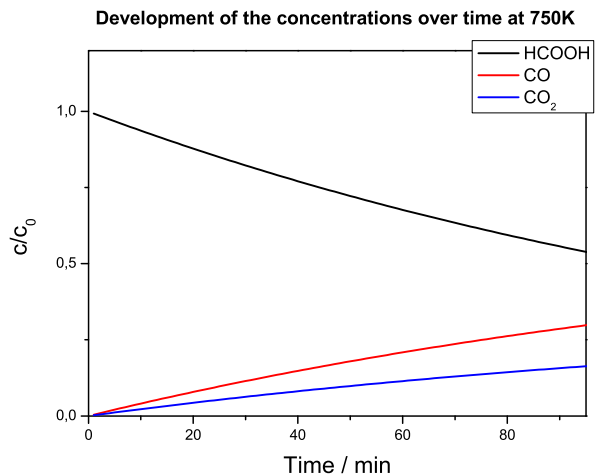}
\caption{Concentrations of HCOOH, CO and CO$_2$ over time at 750K}
\label{concentrations1}
\end{figure}

\newpage

\begin{appendix}

\section{Applying results to reactor conditions} \label{reactor}
In this section we want to apply the results to actual reactor conditions and calculate how long it might take to reduce a copper oxide film.\\
A copper oxide film that measures 1\,nm $\cdot$ 1\,cm $\cdot$ 1\,cm has a volume of $10^{-6}\,\mathrm{cm^3}$. If half of its volume is CuO and half of it is Cu$_2$O then the respective molar amount of oxygen (atoms) is
\begin{align*}
\mathbf{CuO} \qquad &5 \cdot 10^{-7} \mathrm{cm^3} \, \cdot 6.5 \mathrm{\frac{g}{cm^3}} \, \cdot \frac{1}{80 \mathrm{\frac{g}{mol}}} = 4\cdot 10^-8 \mathrm{mol}\\
\mathbf{Cu_{2}O} \qquad &5 \cdot 10^{-7} \mathrm{cm^3} \, \cdot 6 \mathrm{\frac{g}{cm^3}} \, \cdot \frac{1}{143 \mathrm{\frac{g}{mol}}} = 2\cdot 10^-8 \mathrm{mol}\\
\mathbf{total} \qquad & 6\cdot 10^{-8} \mathrm{mol}
\end{align*}
The process runs at 1.3\,mbar and $115\,^{\circ}\mathrm{C}$ (388\,K). Formic acid is supplied with a flow rate of 70\,$\mathrm{\frac{mg}{min}}$ with an Argon carrier gas flow of 100\,sccm (standard cubic cm, approximately 180\,$\mathrm{\frac{mg}{min}}$).\\
We use the ideal gas equation to get the total gas concentration:\\
$$ p V = n R T \qquad \Rightarrow c_{total} = \frac{n}{V} = \frac{p}{R T} = 0.0403 \mathrm{\frac{mol}{m^3}} $$
We compute the partial molar fluxes and the concentration of HCOOH in the reactor,
\begin{align*}
\mathbf{HCOOH} \qquad &\frac{70\mathrm{ \frac{mg}{min} }} {46.03\mathrm{\frac{g}{mol}}} = 1.5207 \cdot 10^{-3} \mathrm{\frac{mol}{min}}\\
\mathbf{Ar} \qquad &\frac{180\mathrm{ \frac{mg}{min} }} {39.95\mathrm{\frac{g}{mol}}} = 4.506 \cdot 10^{-3} \mathrm{\frac{mol}{min}}\\
\mathbf{total} \qquad &6.027 \cdot 10^{-3} \mathrm{\frac{mol}{min}}
\end{align*}
$$c_{HCOOH} = \frac{c_{total}}{4.506 + 1.5207} \cdot 1.5207 = 0.0101688 \frac{mol}{m^3}$$
and the gas flow speed Q.
$$ Q_{tot} = \frac{6.027\mathrm{ \frac{mmol}{min} }} {0.0403\mathrm{\frac{mol}{m^3}}} \mathrm{HCOOH} = 0.1495 \mathrm{\frac{m^3}{min}} = 2.492 \frac{l}{s}$$
Thus, it takes about 0.4 seconds to exchange a litre of the gas. Assuming the gas volume between supply and the sample is just one litre, we can calculate the CO concentration in the gas when it reaches the sample:
\begin{align*}
c_{CO} &= c_{HCOOH} \cdot ( 1 - e^{-k \Delta T} )\\
&= 0.0101688 \frac{mol}{m^3} ( 1 - exp(-2.4\cdot 10^{-21} \mathrm{\frac{1}{s}} \cdot 0.4\mathrm{s}))\\
&= 9.86 \cdot 10^{-24} \mathrm{\frac{mol}{m^3}}
\end{align*}
the CO molar flux is then
$$ Q_{CO} = c_{CO} \cdot Q_{tot} = 9.86 \cdot 10^{-24} \mathrm{\frac{mol}{m^3}} \cdot 0.15 \mathrm{\frac{m^3}{min}} = 1.4786 \mathrm{\frac{mol}{min}}$$
Thus, the time until $6\cdot 10^{-8}$ moles of CO have reached the sample will be about
$$\frac{6\cdot 10^{-8} \mathrm{mol}} {1.48\cdot 10^{-24}\mathrm{\frac{mol}{min}}} \approx 4\cdot 10^{16} \mathrm{min}$$

\end{appendix}

\newpage

\bibliographystyle{unsrt}
\bibliography{library}

\end{document}